\documentclass[paper=a4, 11pt, american]{scrartcl}
\usepackage{times}
\usepackage[margin=2.74cm]{geometry}

\usepackage[T1]{fontenc}
\usepackage[utf8]{inputenc}
\usepackage{babel}
\usepackage{amsmath}
\usepackage{amssymb}
\usepackage{amsthm}
\usepackage{xfrac} 
\usepackage{dsfont}
\usepackage{verbatim}
\usepackage{epigraph} 
\usepackage{graphicx}	

\usepackage[authoryear,round]{natbib}
\usepackage[indentfirst=false,font=normal]{quoting}
\usepackage{latexsym}
\usepackage{authblk} 
\usepackage[unicode=true,bookmarks=true,bookmarksnumbered=false,bookmarksopen=false, breaklinks=false,pdfborder={0 0 0}, backref=false,colorlinks=false]{hyperref}   

\newtheorem*{theorem*}{Theorem}

\begin{document}
\pagenumbering{roman}

\title{The History of \emph{Moral Certainty} as\\
the Pre-History of \emph{Typicality}}

\author{Mario Hubert}

\affil{{\normalsize The American University in Cairo\\
Department of Philosophy}\\
\vspace{0.5cm}
{\normalsize Forthcoming in Bassi, A., Goldstein, S., Tumulka, R., and Zanghì, N., editors, \emph{Physics and the Nature of Reality: Essays in Memory of Detlef Dürr}. Heidelberg: Springer.}\\
\vspace{0.5cm}
\emph{In Memory of my Teacher Detlef Dürr}\\
}

\maketitle
\begin{abstract}
This paper investigates the historical origin and ancestors of typicality, which is now a central concept in Boltzmannian Statistical Mechanics and Bohmian Mechanics. Although Ludwig Boltzmann did not use the word typicality, its main idea, namely, that something happens \emph{almost always} or is valid for \emph{almost all} cases, plays a crucial role for his explanation of how thermodynamic systems approach equilibrium. At the beginning of the 20\textsuperscript{th} century, the focus on \emph{almost always} or \emph{almost everywhere} was fruitful for developing measure theory and probability theory. It was apparently Hugh Everett III who first  mentioned typicality in physics in 1957 while searching for a justification of the Born rule in his interpretation of quantum mechanics. The historically closest concept before these developments is \emph{moral certainty}, which was invented by the medieval French theologian Jean Gerson, and it became a standard concept at least until the Age of Enlightenment, when Jakob Bernoulli proved the Law of Large numbers. 
\end{abstract}

\newpage

\tableofcontents

\newpage

\epigraph{\emph{In the whole conduct of the understanding, there is nothing of more moment than to know when and where, and how far to give assent.}}{--- \textup{John Locke}, Of the Conduct of the Understanding}

\vspace{-0.5cm}
\section{A Brief History of Typicality}
\pagenumbering{arabic}

Typicality\footnote{Detlef Dürr's work on typicality has influenced much of current research of this notion. In his book \emph{Bohmsche Mechanik als Grundlage der Quantenmechanik} written in German, he used the adjective \emph{typisch} or the noun \emph{das Typische} \citep{Durr:2001aa}; later he changed the noun to \emph{Typizität} \citep*{Durr:2017aa}.} is a statistical concept that has been developed in the context of the Boltzmannian approach to Statistical Mechanics and the arrow of time \citep[see, for example,][]{Lebowitz:1993aa,Lebowitz:2008aa,Goldstein:2001aa,Goldstein:2019aa,Bricmont:2022vk} and in the context of Bohmian mechanics \citep{Durr:1992aa,Durr:2009fk,Oldofredi:2016aa}. In Statistical Mechanics, one big problem is to properly reduce the Second Law of Thermodynamics to microphysical processes \citep{Myrvold:2020wp,Robertson:2021wa}. The Boltzmannian approach says that microstates within a specific macrostate \emph{typically} evolve to a macrostate of higher entropy; in other words, \emph{most} or \emph{almost all} microstates within a specific macrostate evolve to a macrostate of higher entropy. In a similar vein, typicality has been used to explain the quantum mechanical Born rule from the microscopic behavior of Bohmian particles \citep*{Durr:1992aa}. 

In both cases, there is a debate about how to  mathematically formalize \emph{almost all} properly  and how explanatorily successful typicality is \citep{Volchan:2007aa,Frigg:2009aa,Frigg:2011ab,Frigg:2012aa,Lazarovici:2015aa}. 
Typicality is usually formalized by a measure on configuration space or phase space. The main use of the measure is to distinguish between ``big'' sets (those with measure close to 1) and ``small'' sets (those with measure close to 0). A problem with using standard measure theory to formalize typicality is that these measures assign sizes to all kinds of sets, also those that are neither ``big'' nor ``small''. One way out would be to consider equivalence classes of absolutely continuous measures or to replace the measure by a new predicate \citep{Maudlin:2019ab}. 

These issues should not bother us here. Instead, my aim is to trace the history of typicality. Where does it come from? How did it evolve? Have there been similar concepts?

The idea of typicality in Statistical Mechanics goes back to the work of Ludwig Boltzmann (1844--1906) on the Second Law of Thermodynamics \citep[see][for a recent exegesis of Boltzmann's original papers]{Darrigol:2018aa}. The concrete problem was to explain under which (microphysical) circumstances a box of gas in non-equilibrium approaches equilibrium. Boltzmann provided this explanation with his $H$-Theorem in 1872. The way he wrote his paper \emph{Weitere Studien über das Wärmegleichgewicht unter Gasmolekülen} (\emph{engl.} Further Studies on Heat Equilibrium among Gas Molecules)\footnote{My translation.}, however, has created some confusion.  Although Boltzmann knew about potential exceptions to the Second Law of Thermodynamics as early as 1868,\footnote{James Clerk Maxwell, whose results Boltzmann developed further, was also aware of those exceptions already in 1867 when he wrote in a letter to his friend Peter Tait about what has become known as \emph{Maxwell's Demon}.} he wrote in 1872 that the $H$ function (the negative of the entropy) ``must necessarily decrease'' \citep[][Sec. 25.2.7]{Darrigol:2013aa}. 

It was only in 1877 that Boltzmann did correct his proof of the $H$-theorem after Loschmidt's reversibility objection in 1876, and so he emphasized that the Second Law of Thermodynamics allows for exceptions considering an appropriate time scale:

\begin{quote}
One cannot prove that for every possible initial positions and velocities of the spheres, their distribution must become more uniform after a very long time; one can only prove that the number of initial states leading to a uniform state is infinitely larger than that of initial states leading to a non-uniform state after a given long time; in the latter case the distribution would again become uniform after an even longer time. (Boltzmann, 1877, quoted in \citealp[][p.\ 775]{Darrigol:2013aa})
\end{quote} 
The reception of Boltzmann's work was already turbulent at his time. Either scholars did not read his work, misunderstood his work, or downplayed his work \citep[Ch.\ 14]{Brush:1976ab}. In any case, from 1877 on, Boltzmann was explicit in his writings and his lectures that a box of gas in non-equilibrium will approach equilibrium for \emph{almost all} initial condition, or, in modern parlance, \emph{typically.} 


Similar ideas of typicality have been applied in mathematics a little earlier in the 19\textsuperscript{th} century.  The German mathematician Carl Friedrich Gauß (1777–1855) and his Belgian colleague Adolphe Quetelet (1796-1874) used the ``most probable value'' in their works on the normal distribution in the period between 1809 and 1857 \citep{Wagner:2020aa}. Being aware of these developments in mathematics, the German sociologist Max Weber (1864--1920) used the word \emph{typical} in his lectures on \emph{General (“Theoretical”) Political Economy} between 1894 and 1898.

In the first half of the 20\textsuperscript{th} century, notions of \emph{almost all}, \emph{almost everywhere}, and \emph{almost always true} were used in the developments of topology, measure theory, probability theory, and classical and quantum statistical mechanics \citep{Plato:1994aa,Hald:1998aa,Goldstein:2010ab}. 



Given these developments, it seems that typicality was first explicitly mentioned in physics by Hugh \citet[][p.\ 460]{Everett:1957aa}, when talking about his \emph{‘Relative State’ Formulation of Quantum Mechanics}, which has later become the many-worlds interpretation of quantum mechanics \citep{Goldstein:2012aa}:
\begin{quotation}
We wish to make quantitative statements about the relative frequencies of the different possible results of observation---which are recorded in the memory---for a typical observer state; but to accomplish this we must have a method for selecting a typical element from a superposition of orthogonal states.  [\dots] The situation here is fully analogous to that of classical Statistical Mechanics, where one puts a measure on trajectories of systems in the phase space by placing a measure on the phase space itself, and then making assertions (such as ergodicity, quasi-ergodicity, etc.) which hold for ``almost all'' trajectories. \citep[][p.\ 460]{Everett:1957aa}
\end{quotation}
This passage is important for two reasons. First, Everett uses the word ``typical'' in passing. Given the uses of similar notions before 1957, Everett seems to address a community that is familiar with the main idea behind typicality, especially those who have worked on ergodicity. Second, with Everett's mentioning of the word ``typical'' he started to unify previous notions of \emph{almost all}, \emph{almost everywhere}, and \emph{almost always true}. This may be a bit of an over-interpretation, but Everett mentions the different uses of \emph{almost all} in Statistical Mechanics, and he wants to use these ideas in quantum mechanics too. 


I think it is not farfetched to set 1877 and 1957 as two historical milestones for the history of typicality, according to current historical knowledge. 1877 is important because Boltzmann explicitly mentions  for the first time in a published paper  that the Second Law of Thermodynamics when reduced to Statistical Mechanics is not a universally valid law and that ``one can only prove that the number of initial states leading to a uniform state is infinitely larger than that of initial states leading to a non-uniform state after a given long time.'' 1957 is important because it is apparently the first time that typicality was used in a physical context. 

 Going back further in time, 1713 is another milestone for the history of typicality, as Jakob Bernoulli's work \emph{Ars Conjectandi} was published in this year, in which he proved the Law of Large Numbers that he based on the notion of \emph{moral certainty}, which is another concept close to typicality. 
The periods 1713--1877  and 1877--1957  pose their own challenges with respect to the historical 
development of typicality: Who worked on similar concepts between 1713--1877, and did they have a particular influence \emph{on} Boltzmann and the physicists working in Statistical Mechanics at the time? Who worked on similar concepts between 1877--1957, and were they influenced \emph{by} Boltzmann and the physicists working in Statistical Mechanics at the time? These questions are largely unanswered up to now, although excellent historical work has been done in the history of probability and measure theory that deals with formalizing the idea of \emph{almost everywhere} in the period after 1877 \citep[see, for instance,][]{Plato:1994aa,Hald:1998aa}.\

In this paper, however, I want to investigate the \emph{pre-history} of typicality, namely the period before 1713. This early period is important not only for historical reasons: 
\begin{enumerate}
\item
The period before 1713 developed essential ideas for subsequent periods. 
\item
We notice a long tradition, in which philosophers and mathematicians tried to get a grip on uncertainty. 
\item
We can understand the essential idea of typicality by investigating its historical precursor \emph{moral certainty}. 
\item
There has been a recent interest to develop in detail a theory of probability that is grounded on typicality \citep*{Durr:2017aa,Maudlin:2019ab,Hubert:2020aa,Allori:2022vg}.\footnote{The fundamental idea to relate typicality with probability has been expressed in \citet*{Durr:1992aa,Goldstein:2001aa,Durr:2009fk,Goldstein:2012aa}.} We can appreciate this research project  by studying how Jakob Bernoulli based his Law of Large Numbers on the notion of moral certainty, which was a standard notion for a couple of centuries before. 
\end{enumerate}

In the following, I will examine in chronological order in which historical situations people came up with ideas similar to typicality. My endpoint will be Bernoulli's \emph{Ars Conjectandi} published posthumously in 1713.


\section{Aristotle: Scientific Demonstrations vs.\ Dialectial Deductions}

Aristotle distinguished two basic kinds of explanations. In the \emph{Posterior Analytics}, he defines a scientific explanation as a demonstration, a deductive proof, from first principles:

\begin{quotation}
By a demonstration I mean a scientific deduction; and by scientific I mean a deduction by possessing which we understand something. If to understand something is what we have posited it to be, then demonstrative understanding in particular must proceed from items which are true and primitive and immediate and more familiar than and prior to and explanatory of the conclusions. \citep[Aristotle, Posterior Analytics, 71b15–25, translated in][p.\ 2–3]{Aristotle:1994aa}
\end{quotation}
The first principles a scientific demonstration relies on are true propositions that have a distinguished epistemic status of being better known than the propositions that logically follow. They are found by induction, which is successful because we have intuition
 (famously stated at the end of the \emph{Posterior Analytics} and also mentioned in the \emph{Nicomachean Ethics}, 1139b19–39, 1140b30–1141a8; \citealp{Bronstein:2012aa,Bronstein:2016aa} for a first-class recent commentary, as well as \citealp[][Ch.\ 4]{Gerson:2009aa}, and \citealp{Salmieri:2014aa}). Scientific demonstrations pose a subclass of deductions. If one can logically deduce from  first principles a  proposition about some state of affairs, one has a scientific explanation and scientific knowledge of these state of affairs, if the deduction elucidates the causes of the explanandum. Standard examples of first principles are universal claims. We can, for example, deduce from \emph{All ravens are birds} and from \emph{All birds have wings} that \emph{All ravens have wings}. These kinds of syllogisms are the archetype of scientific explanations for Aristotle.

Aristotle is aware that scientific demonstrations are a very special group of arguments that require a particularly high standard of accuracy that we can only demand as an ideal\footnote{\citet{Pasnau:2013aa} convincingly argues that epistemology in its entire history from (at least) Aristotle until very recently has been mostly focusing on epistemic ideals. \citet{Hubert:2021uu} endeavor to revive this tradition for a modern theory of understanding \citep[for a meta-epistemological debate, see][]{Carr:2022aa,Thorstad:2023aa}.} in the sciences; therefore, for intellectual training, casual encounters, and philosophical inquiry, we need to lower this high bar (\emph{Topics}, Book 1, Ch. 1--3). In these areas, one needs to replace scientific demonstrations with \emph{dialectial deductions} (Greek: \emph{enthumêma}, another subtype of deductions). Dialectical deductions have premises that are common beliefs (or reputable or probable opinions). As Aristotle says, ``The common beliefs are the things believed by everyone or by most people or by the wise (and among the wise by all or by most or by those most known and commonly recognized)'' \citep[][\emph{Topics}, 100b20--b25]{Aristotle:1995aa}. The conclusions of dialectical deductions do not need to be true but at least sufficiently convincing. 

In the ethical or political realm, Aristotle argues for a precision of arguments that lies between scientific demonstrations and those dialectical deductions:

\begin{quotation}
Therefore in discussing subjects and arguing from evidence, conditioned in this way, we must be satisfied with a broad outline of the truth; that is, in arguing about what is for the most part so from premisses which are for the most part true we must be content to draw conclusions that are similarly qualified. [\dots] for it is a mark of the trained mind never to expect more precision in the treatment of any subject than the nature of that subject permits; for demanding logical demonstrations from a teacher of rhetoric is clearly about as reasonable as accepting mere plausibility from a mathematician. \citep[Aristotle, Nicomachean Ethics, 1094b15–25, translated in][p.\ 5]{Aristotle:2004aa}
\end{quotation}
The premises we use in ethical or political explanations are not universal truths but, at best, true \emph{in most cases}. Therefore, the conclusions we reach from these premises are only true \emph{in most cases} as well.
Asking for universal truths in ethics and politics is not only impossible for all practical purposes, but it would also miss the point of what these disciplines are about in the first place, for ``[i]t is a mark of the trained mind'' to realize what kinds of explanations to expect in a particular context. G.\ W.\ Leibniz echoed Aristotle when he writes in 1670, ``Only that degree of \emph{certainty} is to be had which a given matter admits'' \citep[translated in][p.\ 122]{Leibniz:1989ab}.

There seems to be a shift in the scope of the premises in dialectical deductions as introduced in the \emph{Topics} and  in the \emph{Nicomachean Ethics}. In the \emph{Topics}, the acceptance of the premises is purely epistemic depending on how many and which people support them, and these premises justify the persuasiveness of the conclusion (given the validity of the deduction). This appears to be close to a Bayesian approach, but rather one basing the degree of belief of a proposition on the beliefs of a collective not of a single agent.

In the \emph{Nicomachean Ethics}, on the other hand, the premises are said to be true ``for the most part'', which can be interpreted in (at least) two ways: either in an epistemic (collective Bayesian) way, such that they are regarded as true by most people, or (as I think is more plausible) in an ontological--statistical way, such that the premises are true in most instances or in most situations.
It is this interpretation that captures the central idea of typicality, that something is valid only \emph{in most cases} and not universally valid. But before typicality has been developed in the 20\textsuperscript{th} century, the notion of \emph{moral certainty} elaborated on Aristotle's idea and was a familiar term among philosophers thereafter.

\section{Gerson: The Inventor of Moral Certainty}
Scholastic philosophy combined Christian theology with Ancient Greek philosophy; especially  Aristotle's philosophy played a pivotal role for metaphysics and epistemology. A major project in scholastic philosophy was moral philosophy with the aim of finding rational grounds for how to act morally. The standard approach was to study particular real cases and to extract from them general principles for moral behavior (this method is called \emph{casuistry}, from Latin \emph{casus} meaning ``case'') that are the guiding principles for other (future) occasions. Today, casuistry is still a popular method in ethics, especially in business ethics, bioethics, and ethics of AI. 

A reasonable standard method in casuistry is to find out what the best reasons are for moral behavior. During the Renaissance era, this method was challenged by two skeptical schools \citep[see][for a detailed analysis, which I heavily rely upon in this section]{Schussler:2009aa}: (i) Neo-Pyrrhonism and (ii) Probabilism (\emph{doctrina probabilatis}).

Neo-Pyrrhonism questioned the reliability of weighing different reasons for action against each other aiming for a suspension of judgement \citep{Floridi:2002aa}. The Probabilists were less skeptical and argued that it is possible to act morally despite not knowing the best reasons, as long as sufficient rationality standards were followed. 

Neo-Pyrrhonism and Probabilism noticed that we act and need to act under uncertainty without knowing the best reasons. A problem was that people became too anxious in their actions under these circumstances---this type of exaggerated irrational anxiety was termed \emph{scruples} (Latin: \emph{scrupuli}) at this time.  The Parisian Christian theologian Jean Gerson (1363--1429) tried to solve this dilemma: although it is unrealistic to have infallible knowledge in our decision-making processes, we can aim for sufficient knowledge for our behavior so that we do not need to suffer from scruples, that is, we can have a clear conscience—Gerson, of course, addressed the Medieval Christian community. Building on Aristotle's insight that ``Only that degree of \emph{certainty} is to be had which a given matter admits'' \citep[translated in][p.\ 122]{Leibniz:1989ab}, Gerson coined the term \emph{moral certainty} (Latin: \emph{certitudo moralis}) for the appropriate level of certainty for moral actions that is not deemed sinful and that is not justified to be followed by scruples:

\begin{quote}
There is, however, a moral certainty, which in our purpose is required, and which suffices. And this we have, when, in our recollection and examen of conscience, we find we have done that, which both our own discretion and the good counsel of others suggested, and have for some time been wont commonly so to do. But if our own judgment should not accuse us of mortal sin, then there is no new peril in going to holy communion, $[$\dots$]$ \citep[][pp.\ 57]{Gerson:1883aa}
\end{quote}

There are essentially three paths to reach moral certainty \citep[][p.\ 40]{Gerson:1883aa}: (i) through other people, (ii) through the study of scripture, and (iii) by using your own faculty of reason. Gerson's justification of moral certainty is similar to  Aristotle's justification of common beliefs as the premises in a dialectical deduction: ``The common beliefs are the things believed by everyone or by most people or by the wise (and among the wise by all or by most or by those most known and commonly recognized)'' \citep[][Topics, 100b20--b25]{Aristotle:1995aa}. Aristotle also emphasized the reliance on other people, but, for obvious reasons, he did not add scripture as a means for justification. Reason, for Aristotle, seem to rather play a role for making deductions or for justifying first principles by intuition.  

Gerson continues the above passage and adds important details about moral certainty:

\begin{quote}
$[$\dots$]$ even though, as it may often happen, some slight doubts may come into our mind. These doubts we ought to repel, and we ought to force ourselves to act contrary to them. I call that a slight doubt, when a person judges of a thing, rather that it is just and good, than that it is evil; yet some reasons or thoughts occur to the mind, leading to some hesitation, but still the first judgment appears far the most certain. Now if both sides seem equally probable, we ought to stop till we get more ground for decision one side or other, either by the help of our own reason, or by consultation with others, or by a divine inspiration obtained through prayer. For unless in this mode a person obtain security in himself, he will always judge that he has made a bad confession, and will never feel easy or at peace, and this can never be good. \citep[][pp.\ 57-58]{Gerson:1883aa}
\end{quote}
Moral certainty still leaves room for \emph{slight doubt}, because there always remains a  degree of uncertainty. Still, a good Christian can act upon moral certainty and should not suffer from anxiety that this action is sinful, because everything that could be done and considered has been taken care of. If this person does not understand this, the consequence would be that one remains unhappy and burdened by fear of having done a sinful act, even when one had done everything humanly possible to act morally. Like the Pyrrhonic and Neo-Pyrrhonic Skeptics, Gerson aimed at calmness of the mind (Greek: \emph{ataraxia}). Whereas the Pyrrhonic Skeptics thought to reach this state by suspension of judgement, as for any argument there can be found a counterargument, Gerson suggested to reasonably lower the bar for certainty that is suitable for the average Christian to act morally without suffering from scruples. Political and ethical experts, on the other hand, can reach a higher level of certainty closer to Aristotle's proposed level in the \emph{Nicomachean Ethics} \citep[][p.\ 453]{Schussler:2009aa}. 


After Gerson invented \emph{moral certainty} it has become a standard concept in scholastic philosophy.\footnote{In the 18\textsuperscript{th} century, the concept of moral certainty entered the Anglo-Saxon legal system, and it merged into the concept of \emph{reasonable doubt} by 1824 \citep{Waldman:1959aa}. Other concepts that were used before moral certainty in the judicial system were \emph{satisfied conscience} and \emph{satisfied understanding} \citep[][p.\ 20]{Shapiro:2012aa}.} It seems to have been so prevalent and useful that even stark critics of scholasticism, like René Descartes (1596-1650), used it without any scruples. 

\section{Descartes: Moral Certainty vs.\ Absolute/Metaphysical Certainty}
Descartes used the concept of moral certainty implicitly in his \emph{Discourse on the Method} (1637) and explicitly in \emph{The Principles of Philosophy} (1644)—a detailed analysis of these works with respect to moral certainty can be found in \citet{Ariew:2011aa} and \citet{Samjetsabam:2022aa}. In the spirit of Gerson, Descartes wrote in Part 3 in the \emph{Discourse on the Method}:

\begin{quote}
 Similarly, since in everyday life we must often act without delay, it is a most certain truth that when it is not in our power to discern the truest opinions, we must follow the most probable. Even when no opinions appear more probable than any others, we must still adopt some; and having done so we must then regard them not as doubtful, from a practical point of view, but as most true and certain, on the grounds that the reason which made us adopt them is itself true and certain. \citep[][p.\ 123]{Descartes:1985aa}
\end{quote}
Descartes gives a here a manual to adjust our degree of certainty. The best would be, of course, to find out the truth. Since we need to make quick decisions in our practical life, we often do not have the time to thoroughly examine our thoughts and ideas (and even if we had we may not figure out the truth). In this case, we need to ``follow the most probable.'' But this situation of having a most probable opinion is not guaranteed, and still we need to choose one of the available options. If we did so with good reason, we can be morally certain about the correctness of the opinion (even if Descartes does not explicitly mention moral certainty here). Descartes point is that our acts would be at least epistemically justified when we are morally certain of them. This seems to be a slightly different position from Gerson's who seems to say something stronger, namely, that morally certain acts are morally justified so that we would not suffer scruples. It seems plausible that Descartes thinks along these lines too, but he does not say so explicitly. 

For his theoretical project, on the other hand, Descartes wants to treat any proposition that can be doubted however slightly as absolutely false, as he says so in Part 4 of the \emph{Discourse}:\footnote{Descartes' strong position leads to the following problem (I thank Charles Sebens for pointing this out). If Descartes treats the proposition $p$ as false, he would need to treat the proposition $\neg p$ as false, too, as both can be doubted. It would seem to be more reasonable to interpret Descartes' position as withholding belief in $p$ if $p$ can be doubted.}

\begin{quote}
For a long time I had observed, as noted above [in Part 3], that in practical life it is sometimes necessary to act upon opinions which one knows to be quite uncertain just as if they were indubitable. But since I now wished to devote myself solely to the search for truth, I thought it necessary to do the very opposite and reject as if absolutely false everything in which I could imagine the least doubt, in order to see if I was left believing anything that was entirely indubitable. \citep[][pp.\ 126-127]{Descartes:1985aa}
\end{quote}
Descartes introduces here his \emph{methodological skepticism}, in which he merely \emph{pretends} that certain opinions are false if he can doubt them in the slightest. His goal is to find those opinions that he cannot doubt in the slightest in order to establish a firm foundation for his epistemology and natural philosophy. Descartes is here about to use moral certainty within natural philosophy and so outside the realm of theology and moral philosophy \citep[][Ch.\ 11 §3, mentions that moral certainty was used by the first statisticians by 1662]{Wootton:2015aa}. 

\begin{figure}[ht]
 \centering
 \includegraphics[width=13.5cm]{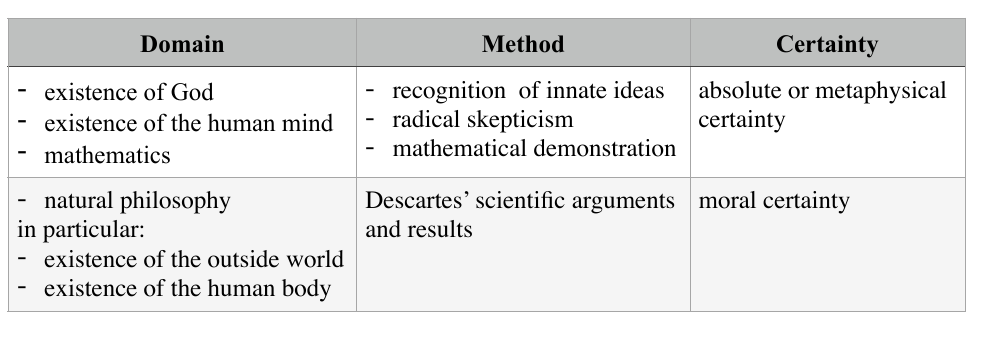}
\caption{Descartes' distinction between absolute and moral certainty. In the \emph{Discourse} and in the \emph{Principles}, Descartes directly argues for the metaphysical or absolute certainty of the existence of God and the existence of the human mind (the human soul). About the existence of the outside world and the human body, Descartes first showed that we can be morally certain; in a second step, he lifts the moral certainty about their existence and even their features to the level of absolute certainty.}
\label{fig:figure-knowledge-Descartes}
\end{figure}

In a subsequent passage in Part 4 of the \emph{Discourse on the Method}, Descartes clarifies his distinction between moral certainty (French: \emph{assurance morale}) and metaphysical certainty (French: \emph{certitude métaphysique}):

\begin{quote}
Finally, if there are still people who are not sufficiently convinced of the existence of God and of their soul by the argument I have proposed, I would have them know that everything else of which they may think themselves more sure—such as their having a body, there being stars and an earth, and the like—is less certain. For although we have a moral certainty about these things, so that it seems we cannot doubt them without being extravagant, nevertheless when it is a question of metaphysical certainty, we cannot reasonably deny that there are adequate grounds for not being entirely sure about them. 
\citep[][pp.\ 129--130]{Descartes:1985aa}
\end{quote}
We have metaphysical certainty of a certain idea or proposition, when we have no rational ground for doubts. Descartes claims that he proved the existence of God and of the human soul to be metaphysically certain. On the other hand,  morally certain claims are less certain, but they are sufficiently certain that they can be only doubted on ``extravagant'' grounds. The existence of the outside world (that humans have a body, that the earth and the stars exist) is morally certain, because we can only doubt them by an extravagant thought experiments, such as a dream (see Fig.\ \ref{fig:figure-knowledge-Descartes}). 

Descartes' use of moral certainty is different from Gerson's in two important respects. First, Descartes applies moral certainty beyond moral philosophy. He is primarily interested in existence claims about the natural world. Although these existence claims are  indeed relevant for our practical actions, they are  rather the preconditions for dealing with the world. Second, Descartes sets a higher bar for moral certainty than Gerson. For all practical purposes, morally certain existence claims about the world can be in principle doubted, but these doubts do not justify us in not believing these claims. Gerson, on the other hand, focuses on morally right behavior which does not require so high a degree of certainty as our belief in an outside world.

At the end of his \emph{Principles of Philosophy}, Descartes examines how certain his philosophical edifice is to be the true story of the world. In \textsection 205 of Part IV, he concludes that his ``explanations appear at least to be morally certain," and he defines moral certainty in the following way:
\begin{quotation}
moral certainty [Latin: \emph{certa moraliter}] is certainty which is sufficient to regulate our behavior, or which measures up to the certainty we have on matters relating to the conduct of life, which we never normally doubt, though we know that it is possible, absolutely speaking, that they may be false. \citep[translated in][p.\ 289]{Descartes:1985aa}
\end{quotation}
Unlike Aristotle, Descartes focuses his distinction on the states of belief of a \emph{single} agent. There are many things in our daily life we have good reasons to take for granted, because they are useful or beneficial for us, although we are not absolutely certain about them \citep[see also][]{Simmons:2001aa}. Even if you have never been to Kuala Lumpur, you are morally certain that this city exists, because of the pictures you have seen and the testimonies of other people, and you would buy a ticket if you wish to visit this city lacking absolute certainty of its existence. Descartes gave a similar example with the city of Rome, and then poses an analogy that his theory is morally certain: Imagine you have an encrypted message and you stumble on a key that is able to decode this message into some sensible text, a string of words that is grammatically correct and meaningful. Then you can be morally certain, certain beyond any reasonable doubt or certain for all practical purposes, that this was the correct key, although it is logically and physically possible that the writer of the message indeed used a different key that would result in a different message. By analogy, Descartes argues, that we can be morally certain about the truth of the axioms that he introduces in the \emph{Principles of Philosophy} because they coherently decode and explain the natural world. Different axioms or principles would be logically possible but highly implausible or unlikely, given the success of his system. 

In this context, moral certainty is related to the \emph{inference to the best explanation} \citep[][]{Lipton:2004aa}. The best explanation that a certain key will decrypt an encrypted message into a sensible text is that this key is the right key. Given the accuracy and success of Descartes' system, it best explains the behavior of the world. It seems that if $X$ is the best explanation for $Y$, we can be more certain about the truth of $X$ than the alternative explanations $X^\prime$. As \citet{Lipton:2000aa} says, ``Inference to the Best Explanation must thus be glossed by the more accurate but less memorable phrase, `inference to the best of the available competing explanations, when the best one is sufficiently good'.''
If $X$ is \emph{by far} the best explanation of $Y$ then we may be morally certain about the truth of $X$.\footnote{I thank Charles Sebens for pointing out the subtleties regarding the relationship between moral certainty and the inference to the best explanation.} Moral certainty is also used in other situations, like in moral behavior, where it does not make sense to talk about a best \emph{explanation}, but rather about the right moral behavior or the best moral behavior. We may call this situation \emph{inference to the best moral behavior}.

In the following paragraph (\textsection 206), Descartes argues that certain aspects of his theory of the world are even \emph{absolutely certain} (Latin: \emph{certa absolute}):
\begin{quotation}
Absolute certainty arises when we believe that it is wholly impossible that something should be otherwise than we judge it to be. This certainty is based on a metaphysical foundation, namely that God is supremely good and in no way a deceiver, and hence that the faculty which he gave us for distinguishing truth from falsehood cannot leas us into error, so long as we are suing it properly and are thereby perceiving something distinctly. Mathematical demonstrations have this kind of certainty, as does the knowledge that material things exist; and the same goes for all evident reasoning about material things. And perhaps even these results of mine will be allowed into the class of absolute certainties, if people consider how they have been deduced in an unbroken chain from first and simple principles of human knowledge. \citep[translated in][p.\ 290]{Descartes:1985aa}
\end{quotation}

While Descartes referred to this degree of certainty in the \emph{Discourse} as \emph{metaphysical certainty}, he changed the terminology in the \emph{Principles} to \emph{absolute certainty}. Apart from the existence of God, the existence of the human soul, and mathematical results, Descartes wants to argue that even some features of his natural philosophy are justified to be true to a higher degree than moral certainty.\footnote{Descartes also strives for the absolute certainty of the outside world in his \emph{Meditations}, which I have left out in this section for lack of space.} He gives two reasons. First, since God is not a deceiver, He created us so that we are able to have epistemic access to the true structure of the world. Second, Descartes deduced his results from true first principles.

\section{Locke: Probability vs.\ Real Certainty}
The period after Descartes, especially in 17\textsuperscript{th} century England, was marked by meticulous refinements of theories explaining and guiding an agent's degree of certainty given a specific domain and specific arguments within the domain \citep[][Ch.\ 2]{Shapiro:1985aa}. Scholars, at this time, largely agreed that demonstrations, like mathematical proofs, lead to the highest level of certainty humans can attain, and they also largely agreed that one cannot demand this high epistemic standard for other disciplines, like the natural sciences, history, law, or even religion. They disagreed, however, on three things: (i) the domain for reaching moral certainty, (ii) the level of certainty for moral certainty, (iii) levels of certainty below moral certainty. 

As different scholars made different proposals for hierarchies of knowledge, John Locke (1632--1704) wrote \emph{An Essay Concerning Human Understanding} in 1689 to synthesize this complex debate, which had a huge influence later on. Although much of Locke scholarship went into analyzing his theory of knowledge, it seems that Locke's aim was to elucidate how one can properly make rational \emph{probable} judgements \citep{Wolterstorff:1996aa,Boespflug:2023aa}. For our purposes, I want to focus on two aspects of Locke's \emph{Essay}: (i) he subsumed absolute and moral certainty within his  notion of \emph{real certainty} (Enquiry, Book IV, Ch. IV), (ii) for probable reasoning, he establishes, what \citet[][p.\ 79]{Wolterstorff:1996aa} calls \emph{The Principle of Proportionality} (Enquiry, Book IV, Ch. XV \& XVI), which is the historical ancestor of David Lewis's  \emph{Principal Principle} \citep{Lewis:1981aa}. 

\begin{figure}[ht]
 \centering
 \includegraphics[width=13.5cm]{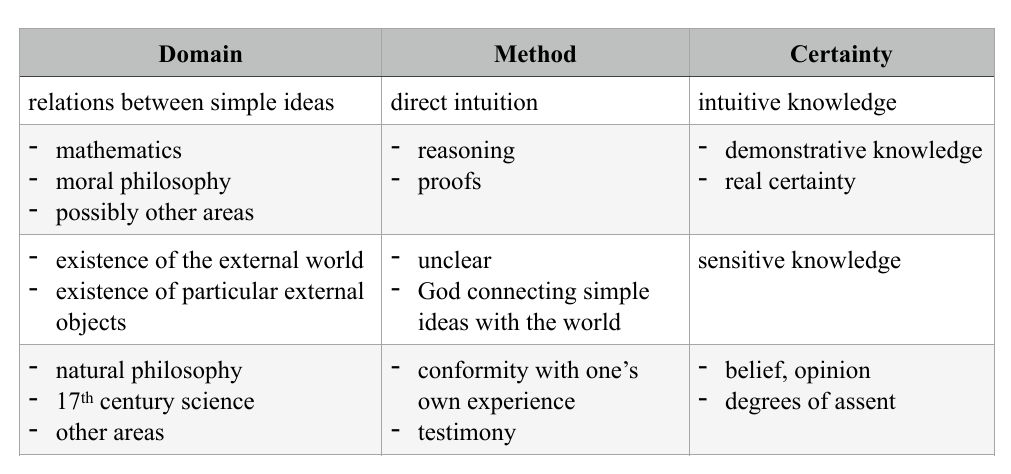}
\caption{Locke's Hierarchies of Knowledge and Probability \citep[][Book IV, Ch. II]{Locke:1689aa}.}
\label{fig:figure-knowledge-Locke}
\end{figure}

Let us first discuss Locke's notion of real certainty. For Locke, the highest form of certainty is attained through knowledge, where he defines knowledge as \emph{the perception of the agreement and disagreement of our own ideas}. Ideas are the contents of our mind, and we have the capacity to compare these ideas. For example, we have an idea of the number 2 and an idea of the number 3. By comparing these ideas of these numbers, we find out that 2 is smaller than 3; therefore, we know that 2 is smaller than 3. Nevertheless, not all types of knowledge lead to the same degree of certainty: thus, Locke introduces a hierarchy of three types of knowledge (see Fig.\ \ref{fig:figure-knowledge-Locke}):
\begin{enumerate}
\item
Intuitive knowledge (Book IV, Ch. II, §1),
\item
Demonstrative knowledge (Book IV, Ch. II, §§2--12),
\item
Sensitive Knowledge (Book IV, Ch. II, §14).
\end{enumerate}

Intuitive knowledge is reached when the mind directly or immediately recognizes the relation between simple ideas, for example, that white is not black. One level down the ladder is demonstrative knowledge, which can be reached once the mind has found a proof to connect two ideas upon reasoning. Mathematics is the prime example for demonstrative knowledge. Another level down the ladder, we find sensitive knowledge. This is the knowledge that there exists an external world with particular external objects beyond our mere ideas in our minds. Locke's arguments for this type of knowledge in Book IV, Ch. II, §14 and Ch. IV, §§2--5 is not clear. His main argument, however, sounds Cartesian, because simple ideas match the external world by the acts of God.\footnote{``\emph{First}, the first are simple ideas, which since the mind, as has been showed, can by no means make to itself, must necessarily be the product of things operating on the mind in a natural way, and producing therein those perceptions which by the wisdom and will of our Maker thy are ordained and adapted to.''  (Book IV, Ch. IV, §4).}  
Another step down the ladder, Locke locates belief, opinion, and different degrees of assent. Among other domains, natural philosophy and 17\textsuperscript{th} century science are notable cases here \citep[see also][Ch.\ 6]{Boespflug:2023aa}, and  testimony from other people become crucial, too, as another method for reaching one's degree of assent. 

Building on his justification for the reality of simple ideas in Book IV, Ch. IV, §§2--5, Locke argues that it follows that mathematical knowledge (through proofs) is \emph{real knowledge} (§6). This argument prepares Locke to break with the tradition of moral certainty---he does not even  mention this concept at all in his \emph{Essay}. Remember that moral certainty was originally used as a degree of certainty for moral claims that one can be certain of for all practical purposes. But for Locke, the domain of morality allows for a higher form of certainty, because it is on a par with mathematics, since moral claims can be \emph{proven}.
 A correct moral proof will lead to \emph{real certainty}, which is the same certainty as attained through mathematical demonstrations (see Fig.\ \ref{fig:figure-knowledge-Locke}):\footnote{Locke used \emph{real certainty} only in this context and did not develop this notion further elsewhere.}

\begin{quote} 
And hence it follows that \emph{moral knowledge} is as \emph{capable of real certainty} as mathematics. For certainty being but the perception of the agreement or disagreement of our ideas, and demonstration nothing but the perception of such agreement, by the intervention of other ideas or mediums; our \emph{moral ideas}, as well as mathematical, being \emph{archetypes} themselves, and so adequate and complete ideas; all the agreement or disagreement which we shall find in them will produce real knowledge, as well as in mathematical figures. \citep[][Book IV, Ch. IV, §7]{Locke:1689aa}
\end{quote}
Locke sounds overtly optimistic about our ability for moral knowledge. In contrast to his predecessors, Locke thought in this passage that it is indeed possible to demonstrate moral propositions similar to mathematical propositions. Later, however, he realized that this is not possible \citep[see][Ch.\ 7, for details]{Boespflug:2023aa}.

Let us now turn to Locke's theory of probable reasoning (last row in Fig.\ \ref{fig:figure-knowledge-Locke}) which will culminate in the \emph{Principle of Proportionality}. First, Locke defines probability in contrast to demonstration:


\begin{quotation}
As demonstration is the showing the agreement, or disagreement of two ideas, by the intervention of one or more proofs, which have a constant, immutable, and visible connexion one with another: so \emph{probability} is nothing but the appearance of such an agreement, or disagreement, by the intervention of proofs, whose connexion is not constant and immutable, or at least is not perceived to be so, but is, or appears for the most part to be so, and is enough to induce the mind to \emph{judge} the proposition to be true, or false, rather than the contrary.
\citeyearpar[][Book IV, Ch.\ XV, \textsection 1]{Locke:1689aa}
\end{quotation}
Whereas demonstrations are logical deductions that generate truths, probability signifies a scheme of reasoning that does not rely on logical entailments but rather on entailments that are valid ``for the most part.'' Locke realizes that these kinds of looser arguments, which are  prevalent in our everyday life, are still strong enough to be believed to a certain degree, as ``[Man] would be often utterly in the dark, and in most of the Actions of his Life, perfectly at a stand, had he nothing to guide him in the absence of clear and certain Knowledge.'' (Book IV, Ch. XIV §1).

 Since Locke aims to give us a hierarchy of probability, he has to tell us along which criteria we can evaluate the probability of a proposition: 

\begin{quote}
\emph{Probability} then, being to supply the defect of our knowledge, and to guide us where that fails, is always conversant about Propositions, whereof we have no certainty, but only some inducements to receive them. The grounds of it are, in short, these \emph{two} following: 

\emph{First}, the conformity of any thing with our own knowledge, observation, and experience. 
\emph{Secondly}, the testimony of others, vouching their observation and experience. \citep[][Book IV, Ch. XV, §4]{Locke:1689aa}
\end{quote}
Locke presents two criteria for evaluating the probability of a proposition: (i) how much the arguments put forward agree (or conform) with our previous knowledge, observation, and experience, (ii) the trustworthiness of the testimony of others.\footnote{Locke warns us to take \emph{the opinions of others} as a criterion for probability, though (Book IV, Ch. XV, §6).} When we evaluate a proposition along these two axes, we can assign it a degree of probability in the following order: (i) demonstration, (ii) improbability, (iii) unlikeliness, (iv) impossibility (Book IV, Ch. XV, §2).
Having done this, the question is now how to adjust one's degree of assent according to this categorization. This is where Locke introduces his \emph{Principle of Proportionality}:

\begin{quote}
\emph{the mind if it will proceed rationally ought to examine all the grounds of probability}, and see how they make more or less, for or against any probable proposition, before it assents to or dissents from it, and upon a due balancing the whole, reject or receive it, with a more or less firm assent, proportionably to the preponderancy of the greater grounds of probability on one side or the other. \citep[][Book IV, Ch. XV, §5]{Locke:1689aa}
\end{quote}
Roughly, the \emph{Principle of Proportionality} says that one should adjust one's assent (or degree of certainty) in accordance to the probability of the proposition.\footnote{This is very close to the \emph{Principal Principle}. A detailed analysis of these two principles would go beyond the scope of this paper.} For the degree of assent, Locke proposes this hierarchy: (i) full assurance and confidence, (ii) conjecture, (iii) doubt, and (iv) distrust.

With the invention of probability calculus, Locke's \emph{Principle of Proportionality} would get  more fine-grained steps for both probability and assent.  Jakob Bernoulli will prove a crucial theorem in probability theory, the Law of Large Numbers, which combines combines probability, assent, and frequencies.

\section{Jakob Bernoulli: Moral Certainty and the Law of Large Numbers}

In this work, Bernoulli introduced and rigorously proved the first limit theorem in mathematics, which Siméon Denis Poisson later called the \emph{Law of Large Numbers} and of which Bernoulli himself said later on, ``I esteem this discovery more than if I had given the quadrature of the circle itself, which even if it were found very great, would be of little use'' \citep[quoted in][p.\ 50]{Sylla:2016aa}. 

Bernoulli's goal was to find out the best way to assign probabilities to certain events. One can determine the probability of a coin toss by examining the physical qualities and the tossing mechanism of the coin toss. If the coin is symmetrical and the tossing mechanism is not prepared in a special way, the coin has probability $\sfrac{1}{2}$ to land heads or tails; this method is in Bernoulli's terminology \emph{a priori}, because it is about dissecting the causes of the phenomenon \citep[][p.\ 327]{Bernoulli:2006aa}.

Bernoulli suggested that it had to be possible to assign the right probabilities also from empirical data, even in cases where no a priori method is possible. This is the \emph{a posteriori} method of assigning probabilities: 
\begin{quotation}
What cannot be ascertained a priori, may at least be found out a posteriori from the results many times observed in similar situations, since it should be presumed that something can happen or not happen in the future in as many cases as it was observed to happen or not to happen in similar circumstances in the past \citep[][p.\ 327]{Bernoulli:2006aa}.
\end{quotation}
Making this a posteriori strategy mathematically rigorous was the aim of the Law of Large Numbers. Consider Bernoulli's example of a large urn consisting of $\sfrac{1}{3}$ black and $\sfrac{2}{3}$ white balls (\citealp[mentioned in a letter to Leibniz, quoted in][p.\ 40]{Bernoulli:2006aa}, also in the \emph{Ars Conjectandi} on p.\ 328)---in this case, $\sfrac{1}{3}$ and $\sfrac{2}{3}$ would be the a priori probabilities. 
The Law of Large Numbers in Bernoulli's interpretation says that, under these circumstances,  you may reach moral certainty after a sufficiently long trial that the true proportion does not deviate too much from $\sfrac{2}{3}$ (more precisely, that it is within $[\sfrac{2}{3}-\epsilon,\sfrac{2}{3}+\epsilon]$ for any given $\epsilon$).

To answer the above question in full mathematical detail, Bernoulli needs a formal definition of moral certainty:

\begin{quotation}
Something is \emph{morally certain} if its probability comes so close to complete certainty that the difference cannot be perceived. By contrast, something is \emph{morally impossible} if it has only as much certainty as the amount by which moral certainty falls short of complete certainty. Thus if we take something that possesses $\sfrac{999}{1000}$ of certainty to be morally certain, then something that has only $\sfrac{1}{1000}$ of certainty will be morally impossible. \citep[][p.\ 316]{Bernoulli:2006aa}
\end{quotation}
Two things are remarkable in this quote. First, Bernoulli introduces precise numbers when to reach moral certainty. Second, these numbers are not universally valid in all circumstances. Moral certainty may have a different mathematical number in different settings \citep[this is also an important feature of the formalization of typicality, see][]{Maudlin:2019ab}. Bernoulli is explicit on that later on:
\begin{quotation}
It would be useful accordingly, if definite limits for moral certainty were established by the authority of the magistracy. For instance, it might be determined whether $\sfrac{99}{100}$ of certainty suffices or whether $\sfrac{999}{1000}$ is required. Then a judge would not be able to favor one side, but would have a reference point to keep constantly in mind in pronouncing a judgment. \citep[][p.\ 321]{Bernoulli:2006aa}
\end{quotation}
There is no unanimously agreed number for what exactly counts as morally certain: in one case $\sfrac{99}{100}$ is sufficient, in another $\sfrac{999}{1000}$ may be required. Bernoulli managed to give a mathematical formalization of moral certainty and in tandem a mathematical theorem, which states under which circumstances one can expect certain empirical results with moral certainty.  For Bernoulli, probabilities are, on the one hand, fractions of certainty, degrees of belief in modern parlance \citep[][p.\ 62]{Sylla:2016aa}; on the other hand, Bernoulli connected these a posteriori probabilities with frequencies for how often one would pick a black or white ball from an urn, for example.

In a modern approach to ground probability on typicality, Bernoulli's picture is turned upside-down: the frequencies are identified with probability, while typicality replaces moral certainty \citep{Durr:2017aa,Maudlin:2019ab,Hubert:2020aa}.  Typicality arises from counting physical degrees of freedom and is therefore by definition not reflecting degrees of belief, but it is \emph{formally} astoundingly similar to Bernoulli's notion of moral certainty. To bridge the gap between counting physical degrees of freedom and certainty, \citet{Wilhelm:2022aa} invokes a new principle of rationality, the \emph{Typical Principle}, to connect typicality claims with belief about these claims. If a proposition makes a typicality claim, an agent should believe this claim (as long as the agent has no further information that would undermine this claim). One cannot be absolutely certain about typicality claims because it is possible that they are violated in a particular case; instead, an agent may be morally certain of them.  



%
%
%
%

\section{Conclusion}

Although moral certainty had a long tradition in philosophy, it is no longer used. With recent works in statistical mechanics and Bohmian mechanics, we have with typicality in fact a notion that is similar to moral certainty. Typicality formalizes the idea of \emph{almost all} by counting physical degrees of freedom by means of a measure, whereas moral certainty is an epistemic notion that we can be \emph{almost certain} about the truth of a proposition. Therefore, we may regard moral certainty as the epistemic counterpart of typicality. 
It requires, however, a deeper analysis to compare both concepts in detail, but my aim here was a historical one: the history of moral certainty can be regarded as the pre-history of typicality.

\section*{Acknowledgements}
I started to write this paper in 2019 when I was a Postdoctoral Instructor at Caltech, and I wish to thank the Division of the Humanities and Social Sciences for the support and resources I have received at the time that made this research possible in the first place. I wish to thank the participants of the \emph{Caltech Probability Seminar} for helpful  discussions. I thank Frederick Eberhardt, Christopher Hitchcock, and Charles Sebens for their insightful comments on a previous draft of the paper. I also thank Ariane Schneck for her helpful comments on the section on Descartes, and I thank Mark Boespflug for his helpful comments on the section on Locke. I wish to thank Barry Loewer, who was the first to tell me about the relation between moral certainty and typicality. I also thank Tim Maudlin for encouraging me to delve into the history of philosophy and for the many hours we spoke about typicality. Lastly, I thank an anonymous referee for helpful comments, especially on clarifying the recent history of typicality. I am very grateful for having been one of Detlef Dürr's students; his teaching and magnanimity have always inspired me.   
\newpage
\bibliographystyle{abbrvnat}
\bibliography{/Users/mariohubert/boxsync/Forschung/Datenbank/references}

\begin{thebibliography}{61}
\providecommand{\natexlab}[1]{#1}
\providecommand{\url}[1]{\texttt{#1}}
\expandafter\ifx\csname urlstyle\endcsname\relax
  \providecommand{\doi}[1]{doi: #1}\else
  \providecommand{\doi}{doi: \begingroup \urlstyle{rm}\Url}\fi

\bibitem[Allori(2022)]{Allori:2022vg}
V.~Allori.
\newblock The paradox of deterministic probabilities.
\newblock \emph{Inquiry}, 2022.
\newblock \doi{10.1080/0020174X.2022.2065530}.

\bibitem[Ariew(2011)]{Ariew:2011aa}
R.~Ariew.
\newblock The new matter theory and its epistemology: Descartes (and late
  scholastics) on hypotheses and moral certainty.
\newblock In P.~Anstey and D.~Jalobeanu, editors, \emph{Vanishing Matter and
  the Laws of Motion: Descartes and Beyond}, Routledge Studies in Seventeenth
  Century Philosophy, chapter~2, pages 31--46. London: Routledge, 2011.

\bibitem[Aristotle(1994)]{Aristotle:1994aa}
Aristotle.
\newblock \emph{Posterior Analytics}.
\newblock Oxford: Clarendon Press, 2\textsuperscript{nd} edition, 1994.
\newblock Translated with a commentary by Jonathan Barnes.

\bibitem[Aristotle(1995)]{Aristotle:1995aa}
Aristotle.
\newblock \emph{Selections}.
\newblock Indianapolis, IN: Hackett Publishing, 1995.
\newblock Translated, with Introduction, Notes, and Glossary by Terence Irwin
  and Gail Fine.

\bibitem[Aristotle(2004)]{Aristotle:2004aa}
Aristotle.
\newblock \emph{The Nicomachean Ethics}.
\newblock London: Penguin, 2004.
\newblock Translated by J.\ A.\ K.\ Thomson.

\bibitem[Bernoulli(2006)]{Bernoulli:2006aa}
J.~Bernoulli.
\newblock \emph{The Art of Conjecturing}.
\newblock Baltimore: The Johns Hopkins University Press, 2006.
\newblock Translated with an introduction and notes by Edith Dudley Sylla.

\bibitem[Boespflug(2023)]{Boespflug:2023aa}
M.~Boespflug.
\newblock \emph{Locke's Twilight of Probability: An Epistemology of Rational
  Assent}.
\newblock New York: Routledge, 2023.

\bibitem[Bricmont(2022)]{Bricmont:2022vk}
J.~Bricmont.
\newblock \emph{Making Sense of Statistical Mechanics}.
\newblock Switzerland: Springer International Publishing, 2022.

\bibitem[Bronstein(2012)]{Bronstein:2012aa}
D.~Bronstein.
\newblock {The Origin and Aim of "Posterior Analytics" II.19}.
\newblock \emph{Phronesis}, 57\penalty0 (1):\penalty0 29--62, 2012.

\bibitem[Bronstein(2016)]{Bronstein:2016aa}
D.~Bronstein.
\newblock \emph{Aristotle on Knowledge and Learning: The \emph{Posterior
  Analytics}}.
\newblock Oxford: Oxford University Press, 2016.

\bibitem[Brush(1976)]{Brush:1976ab}
S.~G. Brush.
\newblock \emph{The Kind of Motion We Call Heat: A History of the Kinetic
  Theory of Gases in the 19th Century}, volume 2 - Statistical Physics and
  Irreversible Processes.
\newblock Amsterdam: North-Holland, 1976.

\bibitem[Carr(2022)]{Carr:2022aa}
J.~Carr.
\newblock Why ideal epistemology?
\newblock \emph{Mind}, 131:\penalty0 1131--1162, 2022.

\bibitem[Darrigol(2018)]{Darrigol:2018aa}
O.~Darrigol.
\newblock \emph{Atoms, Mechanics, and Probability: Ludwig Boltzmann's
  Statistico-Mechanical Writings -- An Exegesis}.
\newblock Oxford: Oxford University Press, 2018.

\bibitem[Darrigol and Renn(2013)]{Darrigol:2013aa}
O.~Darrigol and J.~Renn.
\newblock The emergence of statistical mechanics.
\newblock In J.~Z. Buchwald and R.~Fox, editors, \emph{The Oxford Handbook of
  History of Physics}, chapter~25, pages 765--88. Oxford: Oxford University
  Press, 2013.

\bibitem[Descartes(1985)]{Descartes:1985aa}
R.~Descartes.
\newblock \emph{The Philosophical Writings of Descartes}, volume~1.
\newblock Cambridge, UK: Cambridge University Press, 1985.
\newblock Translated by John Cottingham, Robert Stoothoff, and Dugald Murdoch.

\bibitem[D{\"u}rr(2001)]{Durr:2001aa}
D.~D{\"u}rr.
\newblock \emph{Bohmsche Mechanik als Grundlage der Quantenmechanik}.
\newblock Berlin: Springer, 2001.

\bibitem[D\"urr and Teufel(2009)]{Durr:2009fk}
D.~D\"urr and S.~Teufel.
\newblock \emph{Bohmian Mechanics: The Physics and Mathematics of Quantum
  Theory}.
\newblock Berlin: Springer, 2009.

\bibitem[D\"urr et~al.(1992)D\"urr, Goldstein, and Zangh\`i]{Durr:1992aa}
D.~D\"urr, S.~Goldstein, and N.~Zangh\`i.
\newblock Quantum equilibrium and the origin of absolute uncertainty.
\newblock \emph{Journal of Statistical Physics}, 67\penalty0 (5):\penalty0
  843--907, 1992.

\bibitem[D{\"u}rr et~al.(2017)D{\"u}rr, Froemel, and Kolb]{Durr:2017aa}
D.~D{\"u}rr, A.~Froemel, and M.~Kolb.
\newblock \emph{Einf{\"u}hrung in die {W}ahrscheinlichkeitstheorie als
  {T}heorie der {T}ypizit{\"a}t}.
\newblock Heidelberg: Springer, 2017.

\bibitem[Everett(1957)]{Everett:1957aa}
H.~Everett.
\newblock ``{R}elative state'' formulation of quantum mechanics.
\newblock \emph{Reviews of Modern Physics}, 29\penalty0 (3):\penalty0 454--62,
  1957.

\bibitem[Floridi(2002)]{Floridi:2002aa}
L.~Floridi.
\newblock \emph{Sextus Empiricus: The Transmission and Recovery of Pyrrhonism}.
\newblock Oxford: Oxford University Press, 2002.

\bibitem[Frigg(2009)]{Frigg:2009aa}
R.~Frigg.
\newblock Typicality and the approach to equilibrium in {B}oltzmannian
  statistical mechanics.
\newblock \emph{Philosophy of Science}, 76\penalty0 (5):\penalty0 997--1008,
  2009.

\bibitem[Frigg(2011)]{Frigg:2011ab}
R.~Frigg.
\newblock Why typicality does not explain the approach to equilibrium.
\newblock In M.~Su{\'a}rez, editor, \emph{Probabilities, Causes and
  Propensities in Physics}, chapter~4, pages 77--93. Heidelberg: Springer,
  2011.

\bibitem[Frigg and Werndl(2012)]{Frigg:2012aa}
R.~Frigg and C.~Werndl.
\newblock Demystifying typicality.
\newblock \emph{Philosophy of Science}, 79\penalty0 (5):\penalty0 917--29,
  2012.

\bibitem[Gerson(1883)]{Gerson:1883aa}
J.~Gerson.
\newblock \emph{The snares of devil}.
\newblock London: Thomas Richardson, 1883.
\newblock URL \url{https://archive.org/details/SnaresOfTheDevil/mode/2up}.
\newblock Translated by Beta.

\bibitem[Gerson(2009)]{Gerson:2009aa}
L.~P. Gerson.
\newblock \emph{Ancient Epistemology}.
\newblock Cambridge, UK: Cambridge University Press, 2009.

\bibitem[Goldstein(2001)]{Goldstein:2001aa}
S.~Goldstein.
\newblock Boltzmann's approach to statistical mechanics.
\newblock In J.~Bricmont, D.~D{\"u}rr, M.~Galavotti, G.~Ghirardi,
  F.~Petruccione, and N.~Zangh\`i, editors, \emph{Chance in Physics:
  Foundations and Perspectives}, pages 39--54. Heidelberg: Springer, 2001.

\bibitem[Goldstein(2012)]{Goldstein:2012aa}
S.~Goldstein.
\newblock Typicality and notions of probability in physics.
\newblock In Y.~Ben-Menahem and M.~Hemmo, editors, \emph{Probability in
  Physics}, chapter~4, pages 59--71. Heidelberg: Springer, 2012.

\bibitem[Goldstein et~al.(2010)Goldstein, Lebowitz, Tumulka, and
  Zangh{\`\i}]{Goldstein:2010ab}
S.~Goldstein, L.~J. Lebowitz, R.~Tumulka, and N.~Zangh{\`\i}.
\newblock Long-time behavior of macroscopic quantum systems.
\newblock \emph{The European Physical Journal H}, 35\penalty0 (2):\penalty0
  173--200, 2010.

\bibitem[Goldstein et~al.(2020)Goldstein, Lebowitz, Tumulka, and
  Zangh{\`\i}]{Goldstein:2019aa}
S.~Goldstein, J.~L. Lebowitz, R.~Tumulka, and N.~Zangh{\`\i}.
\newblock Gibbs and boltzmann entropy in classical and quantum mechanics.
\newblock In V.~Allori, editor, \emph{Statistical Mechanics and Scientific
  Explanation: Determinism, Indeterminism and Laws of Nature}, chapter~14,
  pages 519--581. World Scientific, 2020.

\bibitem[Hald(1998)]{Hald:1998aa}
A.~Hald.
\newblock \emph{A History of Mathematical Statistics from 1750 to 1930}.
\newblock Wiley, 1998.

\bibitem[Hubert(2021)]{Hubert:2020aa}
M.~Hubert.
\newblock Reviving frequentism.
\newblock \emph{Synthese}, 199:\penalty0 5255--5284, 2021.

\bibitem[Hubert and Malfatti(2022)]{Hubert:2021uu}
M.~Hubert and F.~Malfatti.
\newblock Towards ideal understanding.
\newblock \emph{Ergo}, 2022.
\newblock Forthcoming.

\bibitem[Lazarovici and Reichert(2015)]{Lazarovici:2015aa}
D.~Lazarovici and P.~Reichert.
\newblock Typicality, irreversibility and the status of macroscopic laws.
\newblock \emph{Erkenntnis}, 80\penalty0 (4):\penalty0 689--716, 2015.

\bibitem[Lebowitz(1993)]{Lebowitz:1993aa}
J.~L. Lebowitz.
\newblock Macroscopic laws, microscopic dynamics, time's arrow and
  {B}oltzmann's entropy.
\newblock \emph{Physica A}, 194:\penalty0 1--27, 1993.

\bibitem[Lebowitz(2008)]{Lebowitz:2008aa}
J.~L. Lebowitz.
\newblock Time's arrow and {B}oltzmann's entropy.
\newblock \emph{Scholarpedia}, 3\penalty0 (4):\penalty0 3348, 2008.

\bibitem[Leibniz(1989)]{Leibniz:1989ab}
G.~W. Leibniz.
\newblock \emph{Philosophical Papers and Letters}.
\newblock Dordrecht: Kluwer, 2\textsuperscript{nd} edition, 1989.
\newblock Translated and edited, with an introduction by Leroy E. Loemker.

\bibitem[Lewis(1981)]{Lewis:1981aa}
D.~Lewis.
\newblock A subjectivist's guide to objective chance.
\newblock In W.~L. Harper, R.~Stalnaker, and G.~Pearce, editors, \emph{IFS:
  Conditionals, Belief, Decision, Chance and Time}, pages 267--97. Springer
  Netherlands, 1981.

\bibitem[Lipton(2000)]{Lipton:2000aa}
P.~Lipton.
\newblock Inference to the best explanation.
\newblock In W.~H. Newton-Smith, editor, \emph{A Companion to the Philosophy of
  Science}, pages 184--193. Oxford: Blackwell, 2000.

\bibitem[Lipton(2004)]{Lipton:2004aa}
P.~Lipton.
\newblock \emph{Inference to the Best Explanation}.
\newblock New York: Routledge, 2004.

\bibitem[Locke(1689/1997)]{Locke:1689aa}
J.~Locke.
\newblock \emph{An Essay Concerning Human Understanding}.
\newblock London: Penguin, 1689/1997.
\newblock Edited by Roger Woolhouse.

\bibitem[Maudlin(2020)]{Maudlin:2019ab}
T.~Maudlin.
\newblock The grammar of typicality.
\newblock In V.~Allori, editor, \emph{Statistical Mechanics and Scientific
  Explanation: Determinism, Indeterminism and Laws of Nature}, chapter~7, pages
  231--51. World Scientific, 2020.

\bibitem[Myrvold(2020)]{Myrvold:2020wp}
W.~C. Myrvold.
\newblock Explaining thermodynamics: What remains to be done?
\newblock In V.~Allori, editor, \emph{Statistical Mechanics and Scientific
  Explanation: Determinism, Indeterminism and Laws of Nature}, chapter~4, pages
  113--143. World Scientific, 2020.

\bibitem[Oldofredi et~al.(2016)Oldofredi, Lazarovici, Deckert, and
  Esfeld]{Oldofredi:2016aa}
A.~Oldofredi, D.~Lazarovici, D.-A. Deckert, and M.~Esfeld.
\newblock From the universe to subsystems: Why quantum mechanics appears more
  stochastic than classical mechanics.
\newblock \emph{Fluctuations and Noise Letters}, 15:\penalty0 1--16, 2016.

\bibitem[Pasnau(2013)]{Pasnau:2013aa}
R.~Pasnau.
\newblock Epistemology idealized.
\newblock \emph{Mind}, 122:\penalty0 987--1021, 2013.

\bibitem[Robertson(2021)]{Robertson:2021wa}
K.~Robertson.
\newblock {In Search of the Holy Grail: How to Reduce the Second Law of
  Thermodynamics}.
\newblock \emph{The British Journal for the Philosophy of Science}, 2021.
\newblock Forthcoming.

\bibitem[Salmieri et~al.(2014)Salmieri, Bronstein, Charles, and
  Lennox]{Salmieri:2014aa}
G.~Salmieri, D.~Bronstein, D.~Charles, and J.~G. Lennox.
\newblock \emph{Episteme}, demonstration, and explanation: A fresh look at
  aristotle's \emph{Posterior Analytics}.
\newblock \emph{Metascience}, 23\penalty0 (1):\penalty0 1--35, 2014.

\bibitem[Samjetsabam(2022)]{Samjetsabam:2022aa}
M.~Samjetsabam.
\newblock Moral certainty of faculty of reason in descartes' discourse.
\newblock \emph{Tattva Journal of Philosophy}, 13\penalty0 (2):\penalty0 1--18,
  2022.
\newblock \doi{10.12726/tjp.26.1}.

\bibitem[Sch{\"u}ssler(2009)]{Schussler:2009aa}
R.~Sch{\"u}ssler.
\newblock {Jean Gerson, Moral Certainty and the Renaissance of Ancient
  Scepticism}.
\newblock \emph{Renaissance Studies}, 23\penalty0 (4):\penalty0 445--462, 2009.
\newblock ISSN 02691213, 14774658.
\newblock URL \url{http://www.jstor.org/stable/24419382}.

\bibitem[Shapiro(1985)]{Shapiro:1985aa}
B.~J. Shapiro.
\newblock \emph{Probability and Certainty in Seventeenth Century England}.
\newblock Princeton, NJ: Princeton University Press, 1985.

\bibitem[Shapiro(2012)]{Shapiro:2012aa}
B.~J. Shapiro.
\newblock Beyond reasonable doubt: The evolution of a concept.
\newblock In Y.~Batsaki, S.~Mukherji, and J.-M. Schramm, editors,
  \emph{Fictions of Knowledge: Fact, Evidence, Doubt}, pages 19--39. Palgrave
  Macmillan UK, London, 2012.
\newblock ISBN 978-0-230-35461-6.
\newblock \doi{10.1057/9780230354616_2}.
\newblock URL \url{https://doi.org/10.1057/9780230354616_2}.

\bibitem[Simmons(2001)]{Simmons:2001aa}
A.~Simmons.
\newblock {Sensible Ends: Latent Teleology in Descartes' Account of Sensation}.
\newblock \emph{Journal of the History of Philosophy}, 39\penalty0
  (1):\penalty0 49--75, 2001.

\bibitem[Sylla(2016)]{Sylla:2016aa}
E.~D. Sylla.
\newblock {Probability in the 17th and 18th Century Continental Europe from the
  Perspective of Jacob Bernoulli's \emph{Art of Conjecturing}}.
\newblock In A.~H{\'a}jek and C.~Hitchcock, editors, \emph{The Oxford Handbook
  of Probability and Philosophy}, chapter~3, pages 50--68. Oxford: Oxford
  University Press, 2016.

\bibitem[Thorstad(2023)]{Thorstad:2023aa}
D.~Thorstad.
\newblock Why bounded rationality (in epistemology)?
\newblock \emph{Philosophy and Phenomenological Research}, 2023.
\newblock Forthcoming.

\bibitem[Volchan(2007)]{Volchan:2007aa}
S.~B. Volchan.
\newblock Probability as typicality.
\newblock \emph{Studies in History and Philosophy of Modern Physics},
  38\penalty0 (4):\penalty0 801--14, 2007.

\bibitem[von Plato(1994)]{Plato:1994aa}
J.~von Plato.
\newblock \emph{Creating Modern Probability: Its Mathematics, Physics and
  Philosophy in Historical Perspective}.
\newblock Cambridge, UK: Cambridge University Press, 1994.

\bibitem[Wagner(2020)]{Wagner:2020aa}
G.~Wagner.
\newblock Typicality and minutis rectis laws: From physics to sociology.
\newblock \emph{Journal for General Philosophy of Science}, 2020.
\newblock \doi{10.1007/s10838-020-09505-7}.

\bibitem[Waldman(1959)]{Waldman:1959aa}
T.~Waldman.
\newblock Origins of the legal doctrine of reasonable doubt.
\newblock \emph{Journal of the History of Ideas}, 20\penalty0 (3):\penalty0
  299--316, 1959.
\newblock ISSN 00225037, 10863222.
\newblock URL \url{http://www.jstor.org/stable/2708111}.

\bibitem[Wilhelm(2022)]{Wilhelm:2022aa}
I.~Wilhelm.
\newblock {The Typical Principle}.
\newblock \emph{The British Journal for the Philosophy of Science}, 2022.
\newblock \doi{10.1086/723240}.
\newblock URL \url{https://doi.org/10.1086/723240}.

\bibitem[Wolterstorff(1996)]{Wolterstorff:1996aa}
N.~Wolterstorff.
\newblock \emph{John Locke and the Ethics of Belief}.
\newblock Cambridge Studies in Religion and Critical Thought. Cambridge, UK:
  Cambridge University Press, 1996.

\bibitem[Wootton(2015)]{Wootton:2015aa}
D.~Wootton.
\newblock \emph{The Invention of Science: A New History of the Scientific
  Revolution}.
\newblock Madrid: HarperCollins Publishers, 2015.

\end{thebibliography}
\end{document}